\documentclass[aps,showpacs,showkeys,superscriptaddress]{revtex4}

\usepackage{graphicx}

\usepackage{comment}

\begin{document}
\title{Electronic properties of phosphorene and graphene nanoribbons with edge vacancies in magnetic field}

\author{J. Smotlacha}\email{smota@centrum.cz}
\affiliation{Bogoliubov Laboratory of Theoretical Physics, Joint
Institute for Nuclear Research, 141980 Dubna, Moscow region, Russia}
\affiliation{Faculty of Nuclear Sciences and Physical Engineering, Czech Technical University, Brehova 7, 110 00 Prague,
Czech Republic}

\author{R. Pincak}\email{pincak@saske.sk}
\affiliation{Bogoliubov Laboratory of Theoretical Physics, Joint
Institute for Nuclear Research, 141980 Dubna, Moscow region, Russia}
\affiliation{Institute of Experimental Physics, Slovak Academy of Sciences,
Watsonova 47,043 53 Kosice, Slovak Republic}

\date{\today}



\begin{abstract}
The graphene and phosphorene nanostructures have a big potential application in a large area of today's research in physics. However, their methods of synthesis still don't allow the production of perfect materials with an intact molecular structure. In this paper, the occurrence of atomic vacancies was considered in the edge structure of the zigzag phosphorene and graphene nanoribbons. For different concentrations of these edge vacancies, their influence on the metallic properties was investigated. The calculations were performed for different sizes of the unit cell. Furthermore, for a smaller size, the influence of a uniform magnetic field was added.
\end{abstract}

\maketitle

\section{Introduction}

The graphene (carbon) nanostructures have been in the center of physical research for more than 10 years. In the last 5 years, the development has been enhanced with additional research on nanostructures based on phosphorus, tin, molybdenum, boron, silicon \cite{lincarvalho,tianguo,licin,lichen}, etc. In this paper, we will be concerned with the electronic properties of the carbon and phosphorus nanostructures.

There is a basic difference between the nanostructures based on carbon and phosphorus: they are $sp^2$- and $sp^3$-hybridized, respectively. As a result, the smooth graphene hexagonal structure is not present in phosphorene, although it is composed of the hexagons as well. Furthermore, phosphorene can exist in 2 configurations: the black phosphorene created by an anisotropic puckered honeycomb lattice and the blue phosphorene. If nothing different follows from the context, the term "phosphorene" will usually denote "black phosphorene" here. It has the most stable crystal structure among several allotropes of phosphorus.
\begin{figure}[htb]		
		 \includegraphics[width=.3\textwidth]{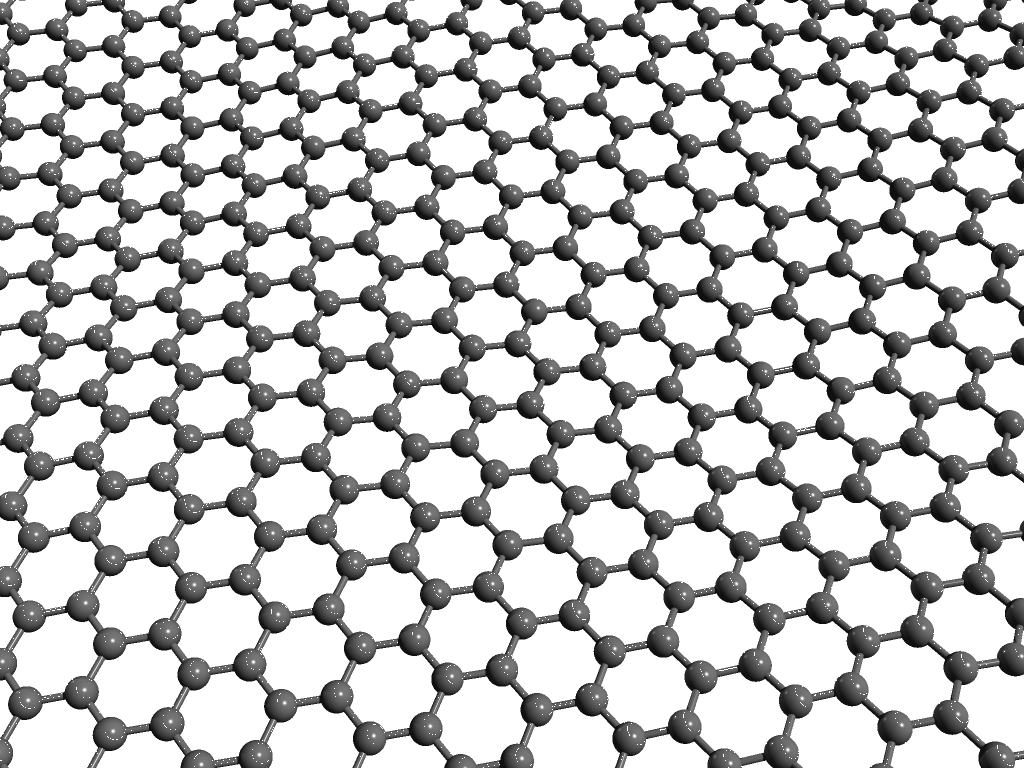}\hspace{2cm}
		        \includegraphics[width=.4\textwidth]{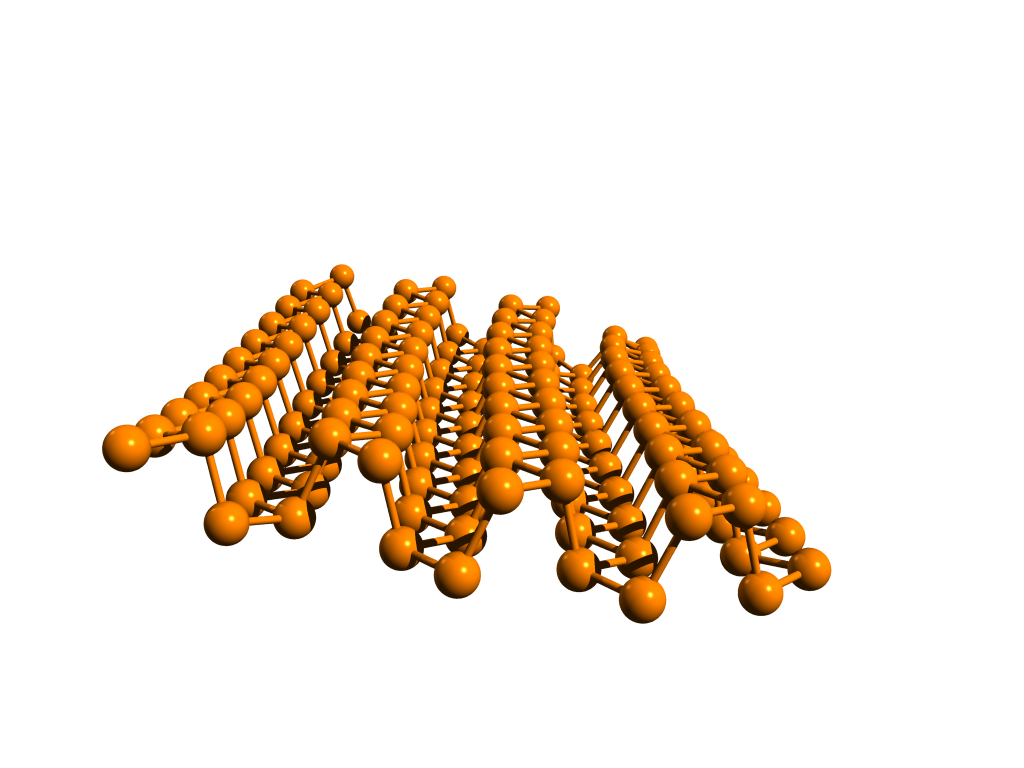}
       		\caption{Molecular structure of graphene (left) and black phosphorene (right).}\label{GNPH}
\end{figure}
Both large-area graphene and phosphorene are sketched in Fig. \ref{GNPH}.

The phosphorene nanostructures are characterized by their energy band gap and high hole mobility \cite{apl1}. The applicability of the black phosphorene as a field effect transistor (FET) is much more significant than that of graphene. On the one hand, the linear dispersion in the corners of the Brillouin zone (BZ) of graphene results in the nearly relativistic velocities of the electrons (the mobility is about $10000\,{\rm cm}^2{\rm V}^{-1}{\rm s}^{-1}$ \cite{schwierz}). On the other hand, the lack of the band gap does not enable one to tune off the graphene channel layer in FET. For the black phosphorene, this problem is canceled, so the corresponding channel in FET can be tuned off. The structural anisotropy is present in the physical properties as well: the hole mobility in the zigzag direction is about $1.8$ times higher than that in the armchair direction \cite{eswaraiah}. It is not so huge as the electron mobility in graphene, but it is still nearly relativistic in the zigzag direction: it is above $1000\,{\rm cm}^2{\rm V}^{-1}{\rm s}^{-1}$ at the temperature 120 K. Similarly to graphene, the band dispersion in the $\Gamma-X$ direction of BZ is linear, unlike the parabolic band dispersion in the $\Gamma-Y$ direction.

The main purpose of this paper is to investigate the influence of the edge vacancies on the electronic properties of the zigzag nanoribbons, i.e., infinitely long strips of constant width that have a characteristic (zigzag) edge structure \cite{wakabayashi1}. Here, they are based on the above mentioned phosphorene (zigzag phosphorene nanoribbons - ZPNR's) and graphene (ZGNR's). Especially, we verify the endurance of the metallic properties (typical for the zigzag nanoribbons) against the Gaussian distribution of the edge vacancies for their different concentrations. Moreover, we investigate a possible influence of the magnetic field on the metallic properties. To calculate the electronic structure, the tight-binding method is used \cite{wallace,slonczewski,nl,iran}.
Other methods based on DFT or the Green function method \cite{nl} can be used in the case of phosphorene. However, the tight-binding method is usual for the graphene-based materials \cite{wallace, wakabayashi1} and for the inclusion of the magnetic field \cite{liu,wakabayashi2}. So it is sufficient for the purpose of comparing the properties of both phosphorene and graphene materials.

As mentioned above, the potential use of the black phosphorene as FET is most effective in the zigzag direction. It evokes an idea to use ZPNR's in FET. But both ZPNR's and ZGNR's are metallic, so the gap for the Fermi energy is missing. However, in ZPNR's, it can be reconstructed with the help of an external electric field \cite{iran}. In this way, the conductance is controlled by the external electric field at Fermi energy which is the transistor effect.

In this paper, after a brief description of the tight-binding method, we calculate the electronic spectrum of ZPNR's and ZGNR's with the atomic vacancies in the edge structure and verify the metallic properties for the Gaussian distribution of the edge vacancies and different sizes of the unit cell. Then, after comparison with the influence of the edge vacancies on the electronic structure of semi-infinite graphene, we will be concerned with the problem of how to improve the metallic properties of the zigzag nanoribbons with a smaller unit cell by the consideration of a uniform magnetic field.\\

\section{Tight-binding method}

\begin{figure}[htb]
\centering
\includegraphics[width=50mm]{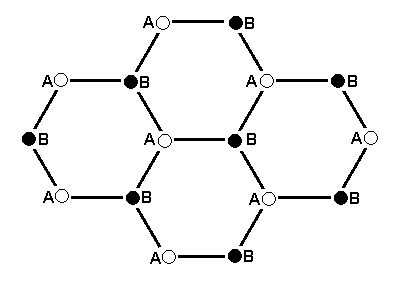}
\hspace{15mm}
\includegraphics[width=0.5\textwidth]{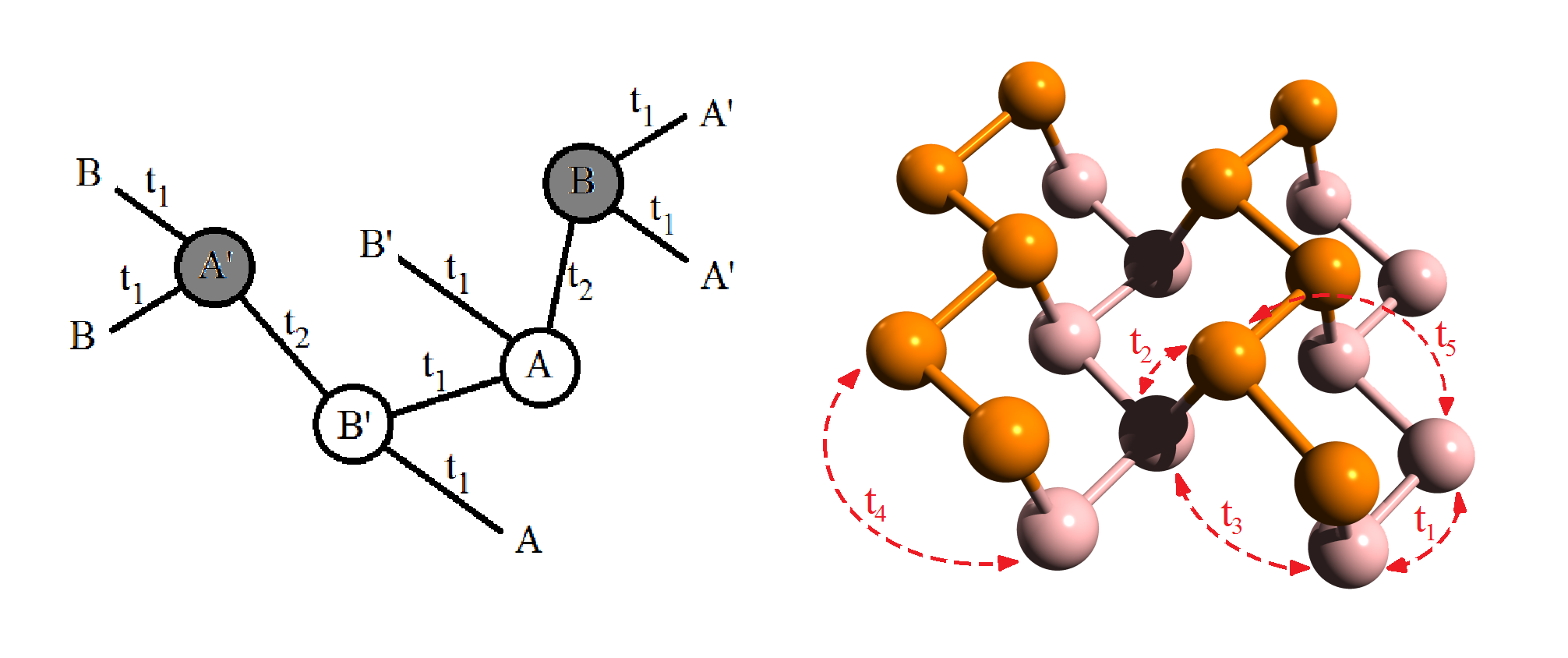}
\caption{Atomic sites and hopping integrals for different periodical structures: graphene (left), black phosphorene (right).}
\label{unit}
\end{figure}

\begin{figure}[htb]		
\includegraphics[width=.17\textwidth]{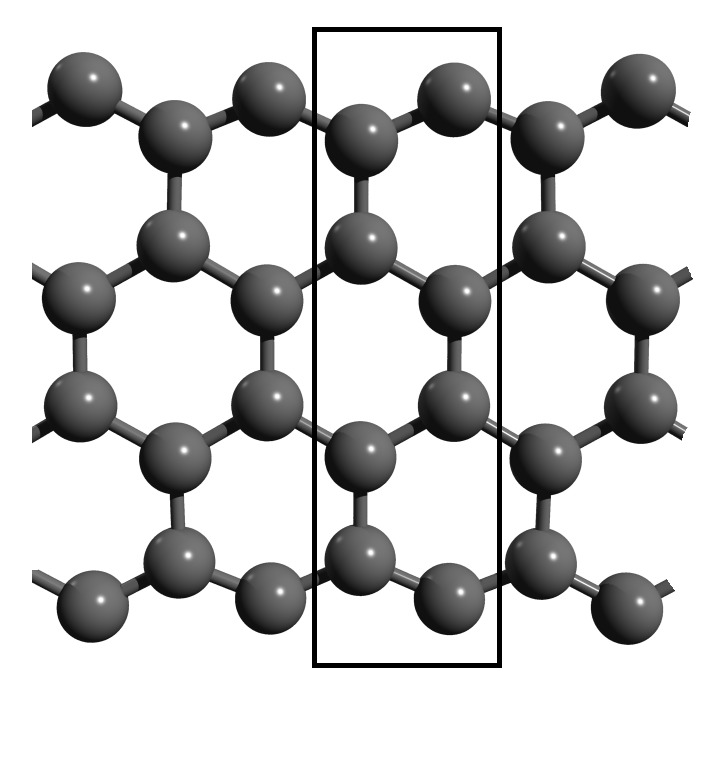}\hspace{10mm}
\includegraphics[width=.3\textwidth]{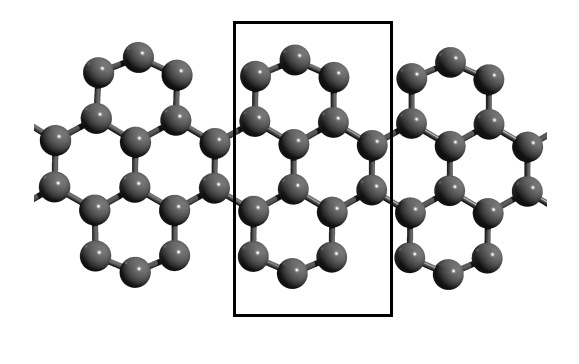}\hspace{10mm}
\includegraphics[width=.4\textwidth]{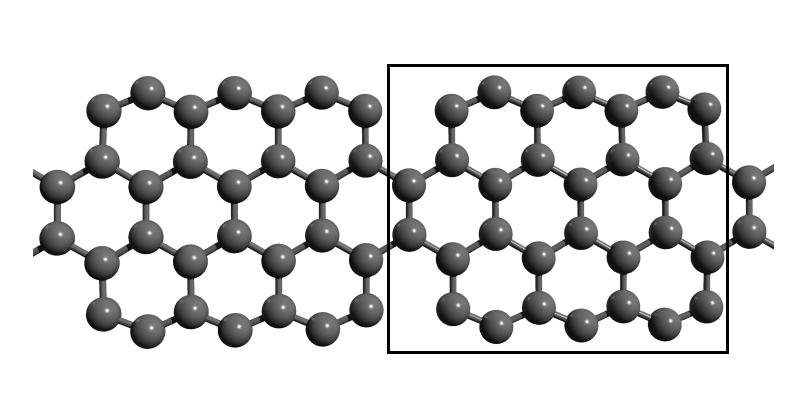}
\caption{Zigzag nanoribbons with different edge structures: no vacancies (left), 1 atom between 2 vacancies (middle), 3 atoms between 2 vacancies (right). The unit cells are denoted by the black frames.}\label{uc}
\end{figure}

In the tight-binding method, the calculation procedure follows from the division of the lattice into inequivalent sublattices. The sublattices are composed of the equivalent atomic sites. In the case of planar graphene or phosphorene, they are denoted by $A,B$ or $A,A',B,B'$, respectively (Fig. \ref{unit}).
The unit cell is determined in this way -- the smallest possible cell containing all the inequivalent atomic sites. Unit cells of different structures are denoted by the black frames in Fig. \ref{uc}.
The interaction between the atoms is characterized by the hopping integrals -- 1 for the graphene structures and 5 for the phosphorene structures \cite{wallace,iran}. For the graphene structures, it is $t=-2.78$ eV, and for the phosphorene structures, we have $t_1=-1.2$ eV, $t_2=3.7$ eV, $t_3=-0.205$ eV, $t_4=-0.105$ eV, and $t_5=-0.055$ eV.

The calculations start on the solution of the Schr\"{o}dinger equation
\begin{equation}\hat{H}\psi=E\psi.\end{equation}
This solution is expressed as the linear combination of the wave functions $\psi_{A_i},\, i=1,...,n$ which correspond to each of $n$ atomic sites $A_i$ in the unit cell. By performing some transformations, we create the matrix elements
\begin{equation}H_{ab}=\int\limits_{\mathcal{R}^3}\psi^{*}_aH\psi_b{\rm d}\overrightarrow{r},\,\,\,a,b\in\{A_1,...,A_n\},\hspace{5mm}
S=\int\limits_{\mathcal{R}^3}\psi^{*}_{A_i}\psi_{A_i}{\rm d}\overrightarrow{r},\,\,\,i=1,...,n.\end{equation}
The resulting electronic spectrum is given by the spectrum of the corresponding matrix \cite{wallace}. By using some additional assumptions \cite{wallace}, in the case of the nanoribbons the resulting matrix equation has the form
\begin{equation}\label{matrixeq2}{\left(\begin{array}{ccccc}
H_{A_1A_1} & H_{A_1A_2} & ... & ... & H_{A_1A_n}\\
H_{A_2A_1} & H_{A_2A_2} & ... & ... & H_{A_2A_n}\\
... & ... & ... & ... & ...\\
... & ... & ... & ... & ...\\
H_{A_nA_1} & H_{A_nA_2} & ... & ... & H_{A_nA_n}\\
\end{array}\right)\left(\begin{array}{c}C_{A_1}\\C_{A_2}\\...\\...\\C_{A_n}
\end{array}\right)=ES\left(\begin{array}{c}C_{A_1}\\C_{A_2}\\...\\...\\C_{A_n}
\end{array}\right),}\end{equation}
where $H_{{A_1}{A_1}}=...=H_{A_nA_n}$. In the case of ZGNR's, the nonzero matrix elements can be written schematically as
\begin{equation}H_{{A_m}{A_n}}=t\Omega_{\vec{k},m,n},\end{equation}
where $\vec{k}$ represents the wave vector. Then, the $j-$th equation of the system has the form
\begin{equation}\label{system2}EC_j=\sum\limits_{l}t\Omega_{\vec{k},j,l}C_l,
\end{equation}
where the index $l$ denotes the nearest  neighboring atomic sites. In the case of ZPNR's whose molecular structure is described by 5 hopping integrals, the number of the terms in the last sum is higher.

Using the results, we can calculate the density of states (DOS) from the definition relation \begin{equation}DOS(E)=\int\limits_{BZ}\delta(E-E(\vec{k})){\rm d}\vec{k}.\end{equation}

\begin{figure}[htb]
\centering
\includegraphics[width=1.2\textwidth]{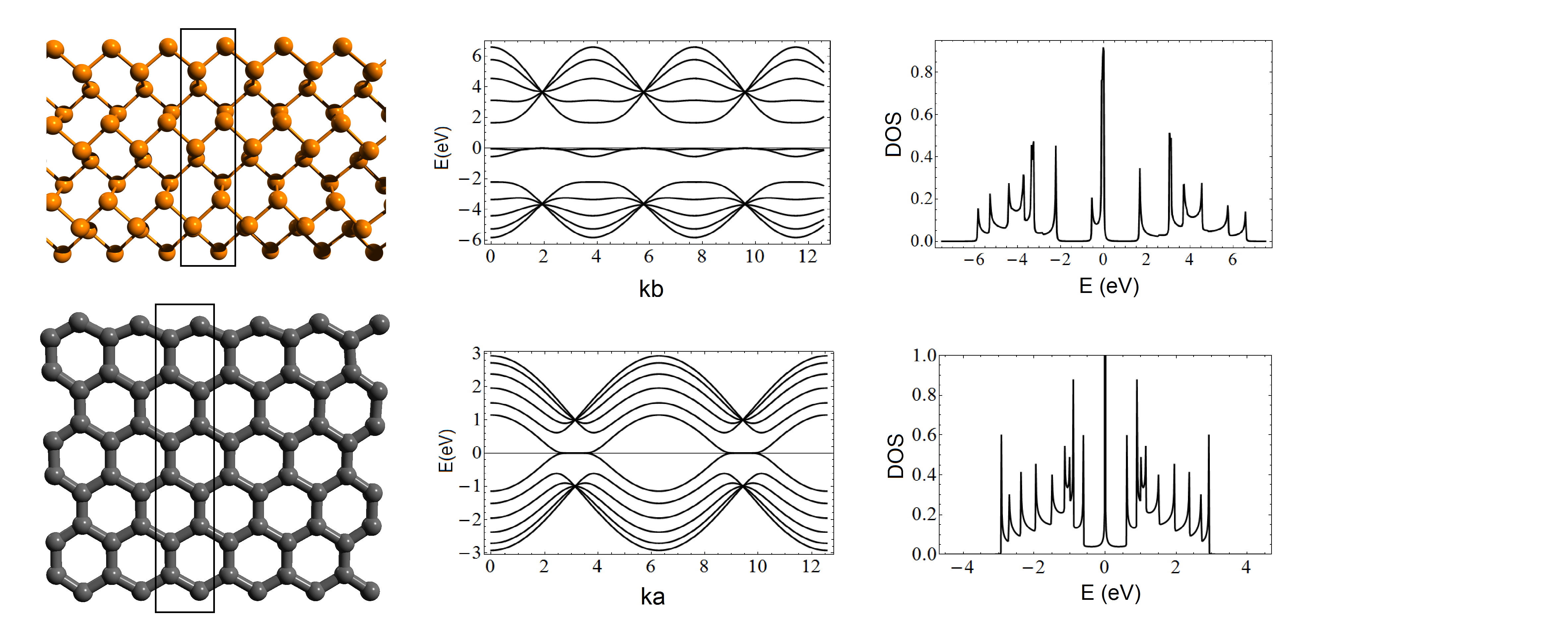}
\caption{Schematic sketch of the zigzag nanoribbon (left), electronic spectrum (middle), and DOS (right) for the case of ZPNR (up) and ZGNR (bottom). The unit cells are denoted by the black frames again.}\label{zznrall}
\end{figure}

In Fig. \ref{zznrall}, we see the comparison of the electronic spectra and DOS of ZPNR and ZGNR \cite{wakabayashi1}. Here and throughout the paper, the zero value of energy corresponds to the Fermi energy. An important (above mentioned) feature typical of zigzag nanoribbons is seen here: DOS contains a characteristic peak at the Fermi energy which indicates the metallic properties. In the electronic spectra, they are demonstrated by a missing gap at the Fermi energy as well.\\

\section{Zigzag nanoribbons with edge vacancies}

Throughout this paper, we restrict our investigations to the zigzag narrow nanoribbons with the width not exceeding $7.5\,{\AA}$ for both ZPNR's and ZGNR's; it means that in the absence of the edge vacancies, their width is created by 8 atoms. In this case, the edge vacancies and the magnetic field have the strongest influence on the electronic structure.
To classify the edge vacancies, we say that 1, 2, 3,... atoms are present between 2 vacancies. This terminology is the shortcut of the fact that 1, 2, 3,... hexagons of the atomic lattice in the edge structure are located between 2 places with the missing atoms (Fig. \ref{uc}).

\begin{figure}[htb]		
		 \includegraphics[width=.45\textwidth]{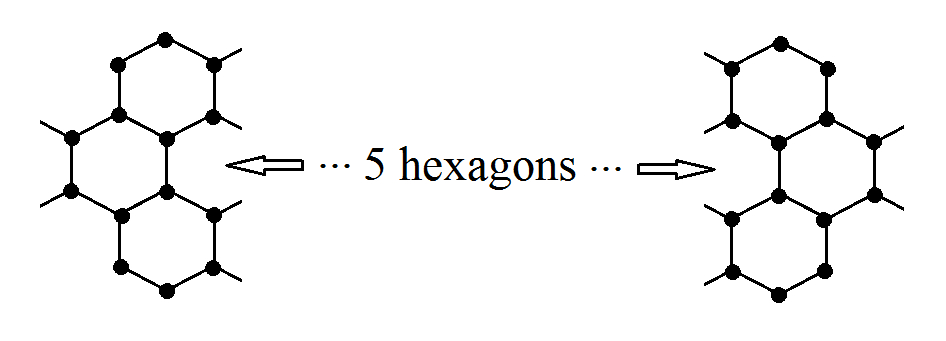}\hspace{5mm}
		\includegraphics[width=.45\textwidth]{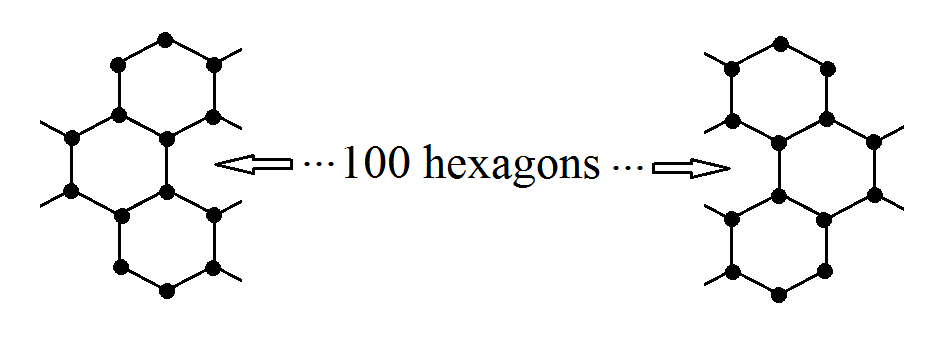}
		\caption{Examples of the unit cells of the investigated zigzag nanoribbons: short unit cell (left), long unit cell(right).}\label{ShortLong}
	\end{figure}

Depending on the external conditions, the distribution of the edge vacancies in the nanoribbons can be uniform or Gaussian. In our calculations, we will suppose the Gaussian distribution of the edge vacancies, which provides a more realistic approach to the real samples. The investigations will be performed for the cases of short (Fig. \ref{ShortLong}, left) as well as long unit cells (Fig. \ref{ShortLong}, right).

\subsection{Small unit cell}\label{csuc}

\begin{figure}[htb]		
\includegraphics[width=.3\textwidth]{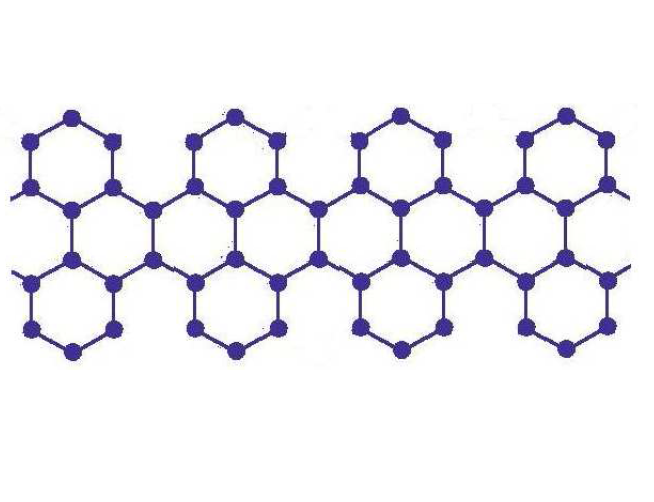}\hspace{3mm}
\includegraphics[width=.3\textwidth]{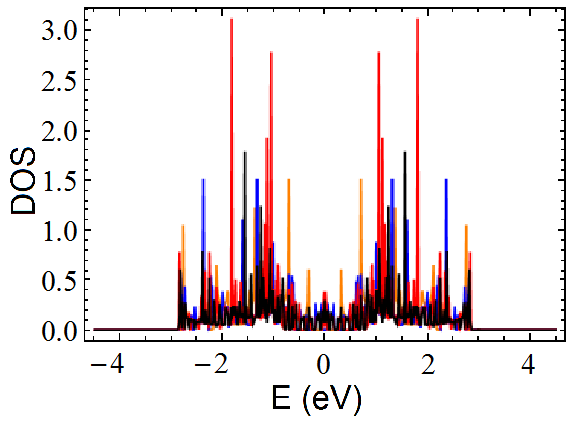}
\includegraphics[width=.33\textwidth]{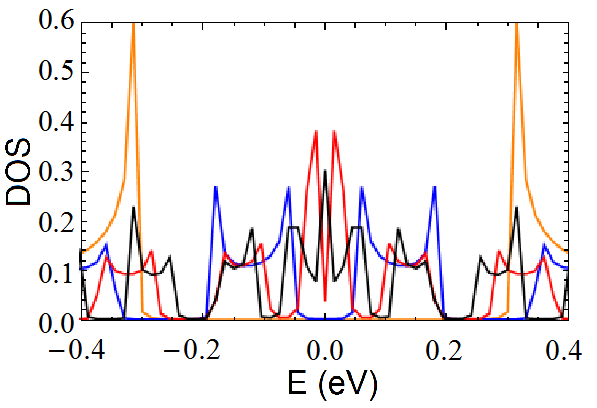}
\caption{DOS of ZGNR's with the edge vacancies for different distances between them at a large scale (middle) and at a small scale (right). There is 1 atom between the vacancies (orange), 5 atoms (blue), 9 atoms (red) and 13 atoms (black). In the left part, we see a sample of ZGNR corresponding to the simplest case of 1 atom between 2 vacancies.}\label{vacLDOS}
\end{figure}

First, we model a test case in which the atomic structures of both edges of the nanoribbon (including the edge vacancies) correspond to each other (Fig. \ref{vacLDOS}, left or the molecular surfaces in Fig. \ref{HB11}). Moreover, we suppose that the placement of the edge vacancies is periodic. Depending on the number of the atoms between 2 vacancies, the length of the unit cell varies -- in this case it is bounded by the vacancies.

\begin{figure}[htb]		
    	        \includegraphics[width=.3\textwidth]{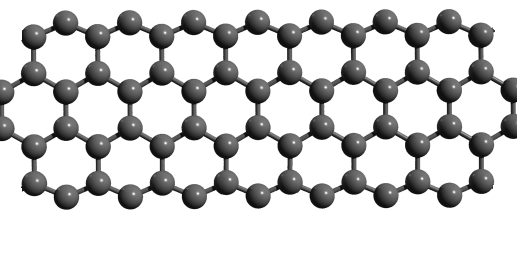}\hspace{5mm}
		        \includegraphics[width=.3\textwidth]{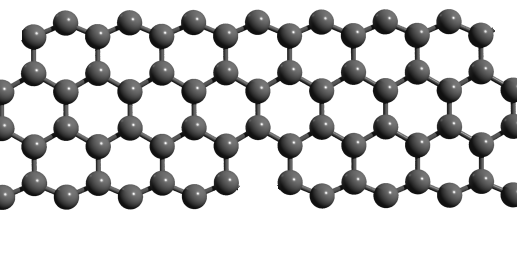}\hspace{5mm}
                \includegraphics[width=.3\textwidth]{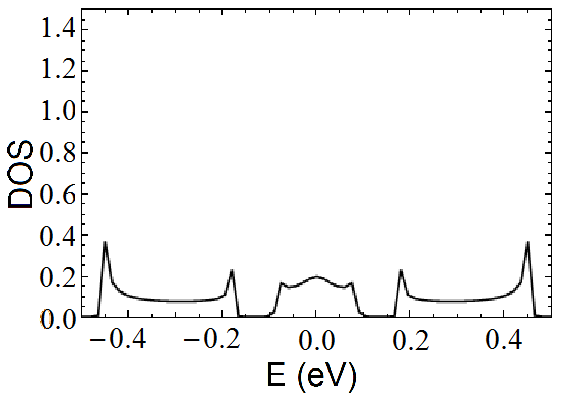}
\caption{Left: structure of ZGNR's, where the vacancies on both edges correspond to each other, in the previous figure we see DOS of this kind of structures; middle: ZGNR with the edge vacancies which are mutually shifted on both edges; right: DOS of ZGNR with the unit cell plotted in the middle of this figure.}\label{vacLDOSa}
\end{figure}

In the case of ZGNR's, the usual peak in DOS vanishes. But it emerges again in DOS of the structures with a longer unit cell. This effect is seen in Fig. \ref{vacLDOS}. It follows from here that the peak in DOS of ZGNR of the given width, which is suppressed for a very short distance between the edge vacancies, emerges again when at least 13 atoms between 2 vacancies are present. (When the width of the nanoribbon is larger than $7.5\,{\AA}$, the number of the atoms between 2 vacancies needed for the recovery of the peak in DOS lowers with the increasing width to 8, 6, 4 etc. Let us note that if the vacancies on both edges are mutually shifted, the peak - or an indication of a peak - in DOS is restored much earlier than in the mentioned case of 13 atoms between 2 vacancies -- see Fig. \ref{vacLDOSa}.)

\begin{figure}[htb]		
\includegraphics[width=.28\textwidth]{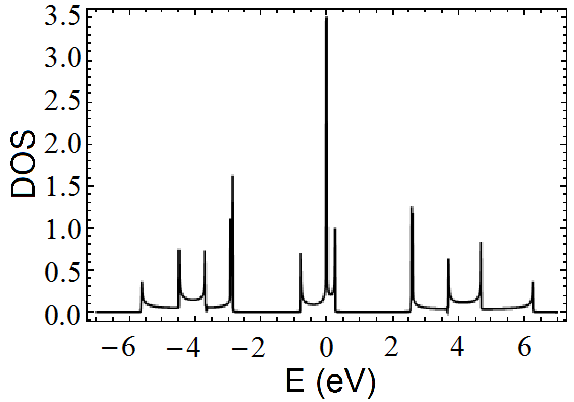}\hspace{1cm}
\includegraphics[width=.27\textwidth]{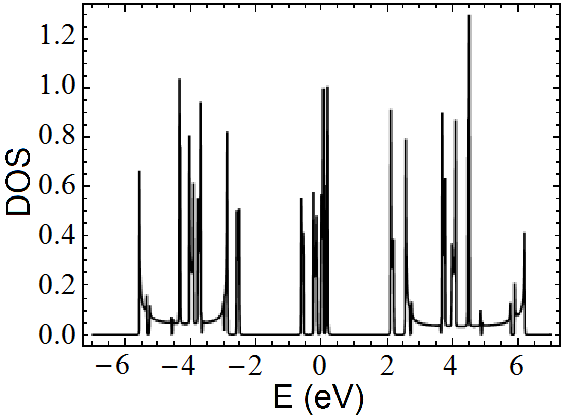}\hspace{1cm}
\includegraphics[width=.28\textwidth]{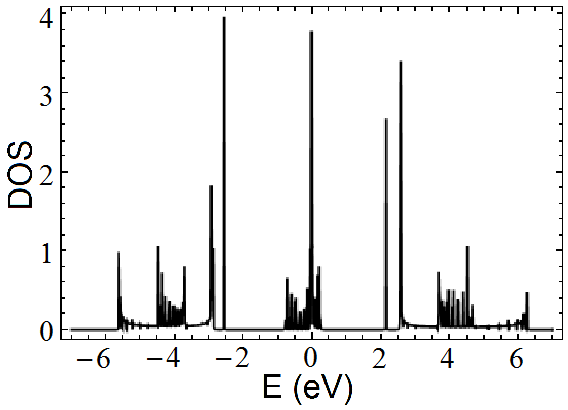}
\caption{DOS of ZPNR's with the edge vacancies for different distances between them. There is no atom between the vacancies (left), 2 atoms (middle) and 9 atoms (right).}\label{vacDOS-ph}
\end{figure}

For ZPNR's, the peak at the Fermi energy in DOS does not vanish for an arbitrary distance between the edge vacancies. Moreover, the positions of the gaps remain. On the other hand, if the size of the unit cell increases, the number of the energy peaks in DOS increases as well. Consequently, the new energy peaks are concentrated around the positions of the energy peaks corresponding to a shorter size of the unit cell. So, in these places, the density of the peaks increases. The amplitude of the peaks around the zero energy area remains, more or less, the same (Fig. \ref{vacDOS-ph}).

\subsection{Large unit cell}

\begin{figure}[htb]		
\includegraphics[width=.35\textwidth]{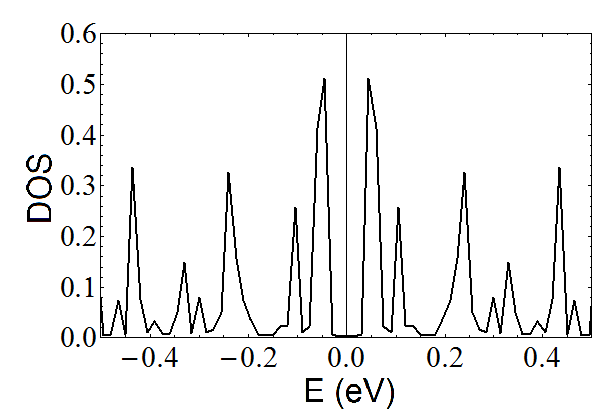}\hspace{1cm}
\includegraphics[width=.35\textwidth]{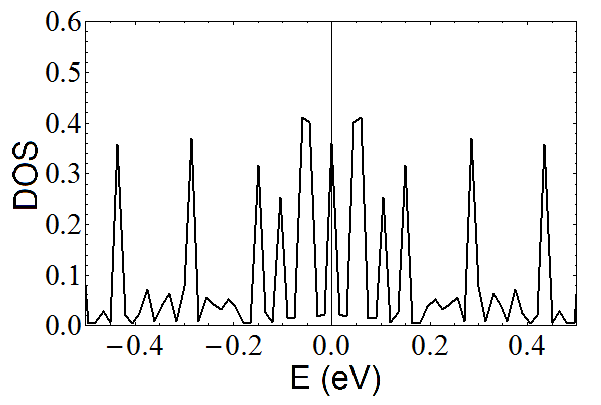}
\caption{DOS of 2 different ZGNR's with the unit cell containing 100 hexagons on one side of the edge structure. This unit cell is composed of the symmetric small subcells. In the case on the left side, the number of the hexagons on one side of the edge structure of these subcells does not exceed 6 hexagons, while in the case on the right side, it achieves 20 hexagons.}\label{pertest}
\end{figure}

Now, we perform the calculation for ZGNR with a significantly larger unit cell by choosing a structure with 100 hexagons in one edge. We consider it to be composed of small unit subcells with the properties specified at the beginning of the previous subsection, i.e., the edge vacancies on both sides mutually correspond. Moreover, we suppose that each of these unit subcells does not contain more than 12 hexagons in the edge structure. In the agreement with the results in \cite{artem}, we can suppose that in this case, the corresponding DOS does not contain any peak at the Fermi level -- this property of each of the subcells is projected into the properties of the whole structure. On the other hand, if at least one of the subcells contains more than 12 hexagons (so, its DOS contains a peak at the Fermi level), a peak at the Fermi energy emerges. This is illustrated in Fig. \ref{pertest}.\\

\subsubsection{Gaussian distribution of the edge vacancies}

In the real conditions, the edge vacancies on both sides of the nanoribbons don't mutually coincide and the most probable distribution of the edge vacancies is Gaussian. For our purpose, we suppose that the Gaussian distribution is applied not to the whole structure, but to each of its unit cells (each of them has the same atomic structure). Similarly to \cite{artem}, we calculate DOS for the structures whose unit cells contain 10, 50, 100, and 400 hexagons in one edge. The considered concentration of the edge vacancies is $30$, $50$, $70$, and $90\%$. For each length of the unit cell (except the last case of 400 hexagons), the presented results are average of the results for 10 different configurations of the edge vacancies.

\begin{figure}[htb]		
    	        \includegraphics[width=.4\textwidth]{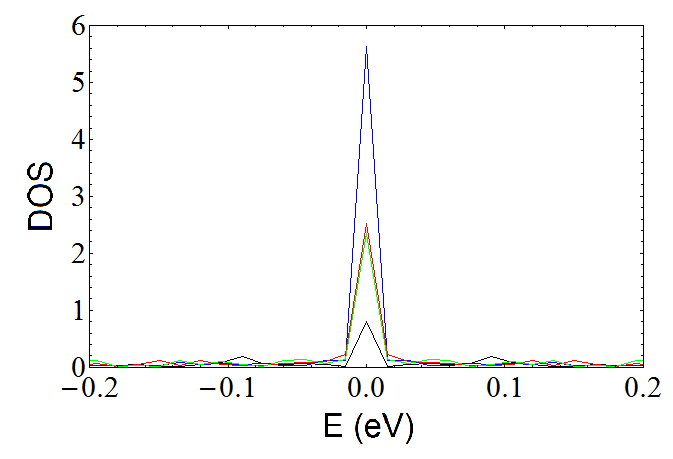}\hspace{1cm}
                \includegraphics[width=.4\textwidth]{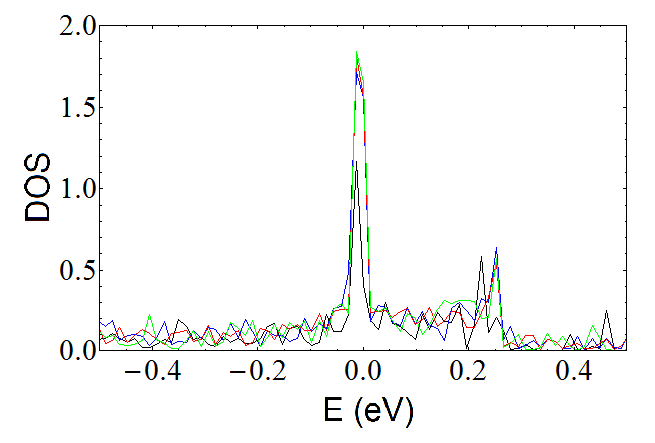}
		       \caption{DOS of ZGNR's (left) and ZPNR's (right) with the Gaussian distribution of the edge vacancies and the concentration $30\%$. The edge structure on one side of the unit cell is created by 10 hexagons (black), 50 hexagons (blue), 100 hexagons (red), 400 hexagons (green).}\label{vacDOSg}
\end{figure}

In Fig. \ref{vacDOSg}, we compare DOS of ZGNR's and ZPNR's with the concentration of the edge vacancies $30\%$. In all the studied cases, the peak at the Fermi energy does not vanish -- there are only some fluctuations in the amplitude in the case of ZGNR with a smaller unit cell. The reason consists in the effect showed in Fig. \ref{vacLDOS}: for a smaller unit cell, the peak at the Fermi energy can vanish for some configurations of the edge vacancies. Then, in the case of a small unit cell, we can't give any reasonable estimate for the amplitude. On the other hand, in the case of ZPNR's, the amplitude seems not to depend very much on the size of the unit cell.

\begin{figure}[htb]		
    	        \includegraphics[width=.4\textwidth]{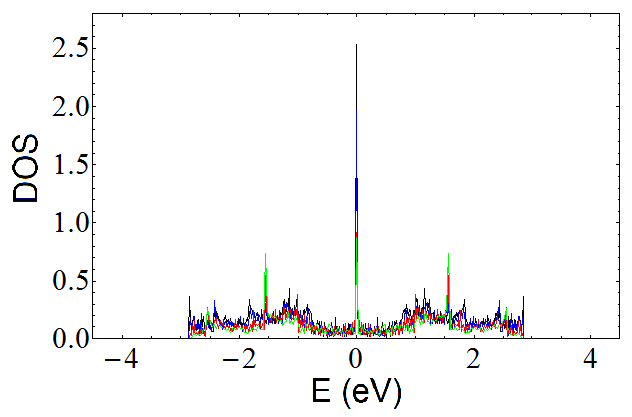}\hspace{1.5cm}
                \includegraphics[width=.4\textwidth]{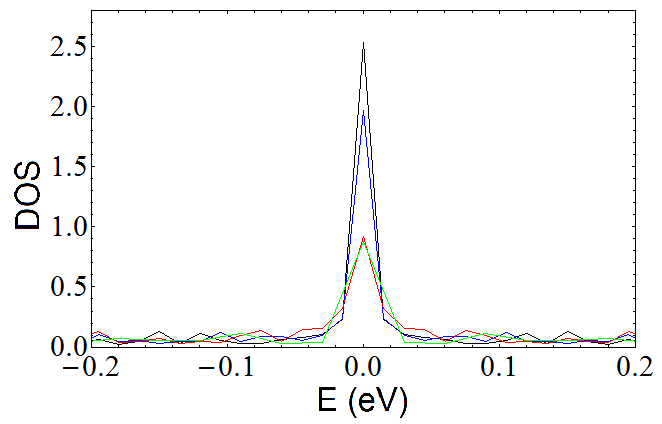}
		       \caption{DOS of ZGNR's with the Gaussian distribution of the edge vacancies on a large and on a small scale. Different concentrations of the edge vacancies are distinguished by colors: $30\%$ -- black, $50\%$ -- blue, $70\%$ -- red, $90\%$ -- green. For the increasing concentration of the edge vacancies, we see a decreasing value of the amplitude of the peak at the Fermi energy.}\label{vacDOSg-gr2}
\end{figure}

\begin{figure}[htb]		
    	        \includegraphics[width=.4\textwidth]{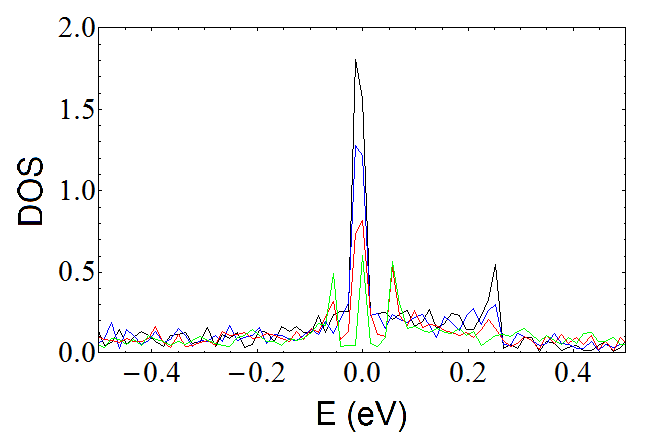}
		       \caption{DOS of ZPNR's with the Gaussian distribution of the edge vacancies. The unit cell is created by 100 hexagons and the concentration of the edge vacancies is: $30\%$ (black), $50\%$ (blue), $70\%$ (red), and $90\%$ (green).}\label{vacDOSg-ph2}
\end{figure}

In Figs. \ref{vacDOSg-gr2} and \ref{vacDOSg-ph2}, we compare the behavior of the amplitudes of the peak at the Fermi energy for both ZGNR's and ZPNR's in the case when one edge of the unit cell is created by 100 hexagons. On the whole, for both ZGNR's and ZPNR's, the amplitude decreases with increasing concentration of the edge vacancies, but it does not vanish.\\

\subsubsection{Comparison with semi-infinite graphene}

We can compare our results with \cite{artem}, where similar calculations were performed for semi-infinite graphene with zigzag edge structure. It can be considered as a zigzag nanoribbon with "semi-infinite" width. It means that compared to our nanoribbon, it has one edge only. This indicates the possibility of slightly different results.

Among others, in \cite{artem} the influence of single edge vacancies on DOS of the whole structure is examined for 3 different distributions: periodic, Gaussian and uniform. The first possibility is investigated in subsection \ref{csuc} of this paper as well (Fig. \ref{vacLDOS}): it is supposed here that the positions of the vacancies on both edges mutually correspond. Then, there must be at least 13 hexagons between 2 edge vacancies to restore the metallic properties. This number is lowered if the vacancies on both edges are mutually shifted (Fig. \ref{vacLDOSa}) or if the width of the structure is increased. In \cite{artem}, the results for the periodic distribution are presented in Fig. 7a. It is demonstrated here that 3 hexagons between 2 edge vacancies are sufficient.

The Gaussian distribution is examined in Fig. 5 of \cite{artem}. Similarly as in the previous subsection, the peak for zero energy does not vanish for the increasing concentration of the edge vacancies, although its amplitude is weakening.

\begin{figure}[htb]		
    	        \includegraphics[width=.4\textwidth]{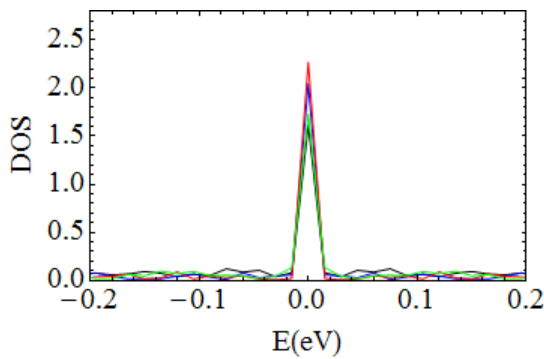}
		       \caption{DOS of ZGNR's with the uniform distribution of the edge vacancies. The unit cell is created by 100 hexagons and the concentration of the edge vacancies is: $30\%$ (black), $50\%$ (blue), $70\%$ (red), and $90\%$ (green).}\label{vacDOSu}
\end{figure}

So, it appears that both papers demonstrate similar influence of the edge vacancies on DOS of the appropriate structures. But one possibility has not yet been explored in this paper - the uniform distribution of the edge vacancies. For this case, the results we see in Fig. \ref{vacDOSu}. It follows from here that this eventuality does not affect the metallic properties very much, moreover, the amplitude of the peak for zero energy first increases with the increasing concentration of the edge vacancies. Only for very high concentrations, it starts to decrease. This behaviour strongly differs from \cite{artem} - Fig. 6 of that paper shows that for semi-infinite graphene and uniform distribution of the edge vacancies, the peak for zero energy disappears for the concentration of the edge vacancies $50\%$.\\

\section{Influence of the magnetic field}

As was demonstrated in the previous section (Figs. \ref{vacLDOS} and \ref{vacLDOSa}), the metallic properties of ZGNR's are suppressed when the distance between the edge vacancies is lower than 13 atoms. One of the possibilities how to weaken this undesirable effect could be switching on a uniform magnetic field. The direction of this magnetic field will be considered perpendicular to the molecular surface. Another possibility would be the in-plane magnetic field which would have zero magnetic flux through the flat surface of ZGNR; so it does not influence the electronic structure and we don't consider it.

Although ZPNR's don't show the loss of the metallic properties in the case of a small unit cell, we will study their behavior under the influence of the magnetic field as well to explore closer the feature found in subsection \ref{csuc} -- the stability of the energy gaps against the changing size of the unit cell. For simplification, we will consider the same (perpendicular) direction of the magnetic field, although due to its geometry, this material does not show the inertion to the in-plane magnetic field (unlike graphene) \cite{FTO}.

\begin{figure}[htb]
\includegraphics[width=0.5\textwidth]{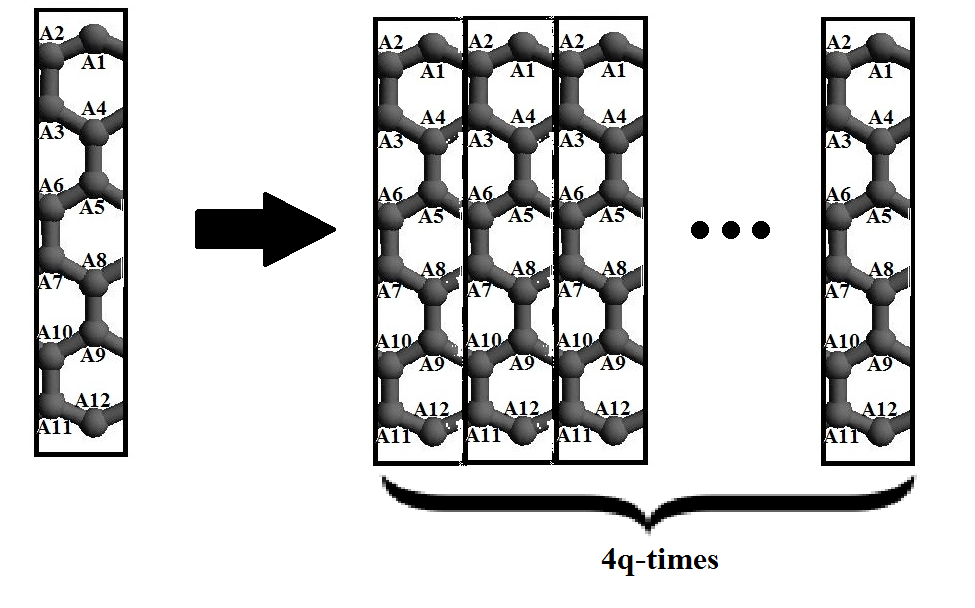}
\caption{Enlargement of the unit cell of a nanoribbon in the presence of a uniform magnetic field.}\label{mcell}
\end{figure}

Under the influence of the magnetic field, the elements of the matrix in (\ref{matrixeq2}) are multiplied by the exponentials depending on the magnetic phase factor and the system of equations (\ref{system2}) can be rewritten into the form
\begin{equation}EC_j=\sum\limits_{l}t\Omega_{\vec{k},j,l}C_l\,\rightarrow\,EC_j=
\sum\limits_{l}t\exp({\rm i}\gamma_{jl})\Omega_{\vec{k},j,l}C_l.\end{equation}
The resulting system of equations is called the Harper equations \cite{harper,liu}. Here, $\gamma_{jl}$ is the magnetic phase factor. It is proportional to $\Phi$ -- the magnetic flux through the unit cell. Let $\Phi_0$ be the magnetic flux quantum. If $\Phi/\Phi_0=p/q$ with $p,q$ mutually primes, then the magnetic phase factor has the period $4q$. As a consequence, the size of the unit cell is enlarged $4q$-times (Fig. \ref{mcell}) and in the same way, the size of the Hamiltonian matrix is changed.

As we are interested in the influence of the magnetic field on the electronic spectrum, it will be investigated depending not on the wave vector but on the ratio $f=\Phi/\Phi_0$ which is the magnetic flux in the units of the magnetic flux quantum.

\subsection{Zigzag nanoribbons without edge vacancies}

\begin{figure}[htb]		
		 \includegraphics[width=.4\textwidth]{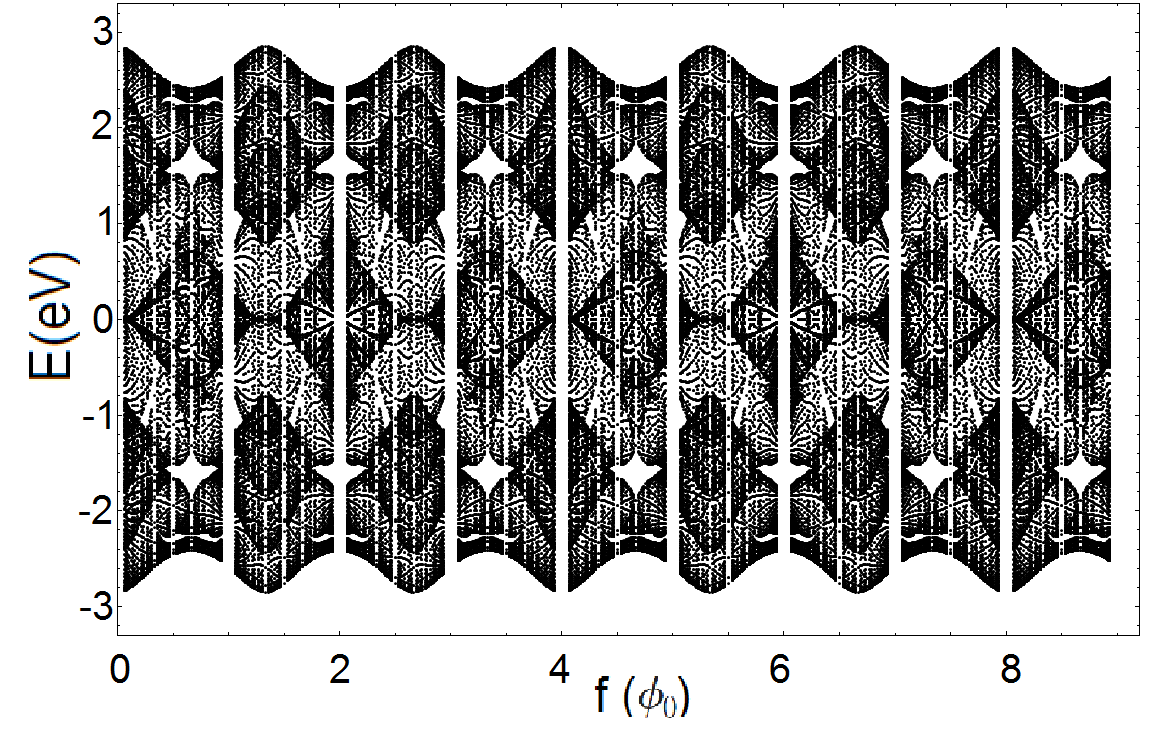}\hspace{15mm}
		 \includegraphics[width=.4\textwidth]{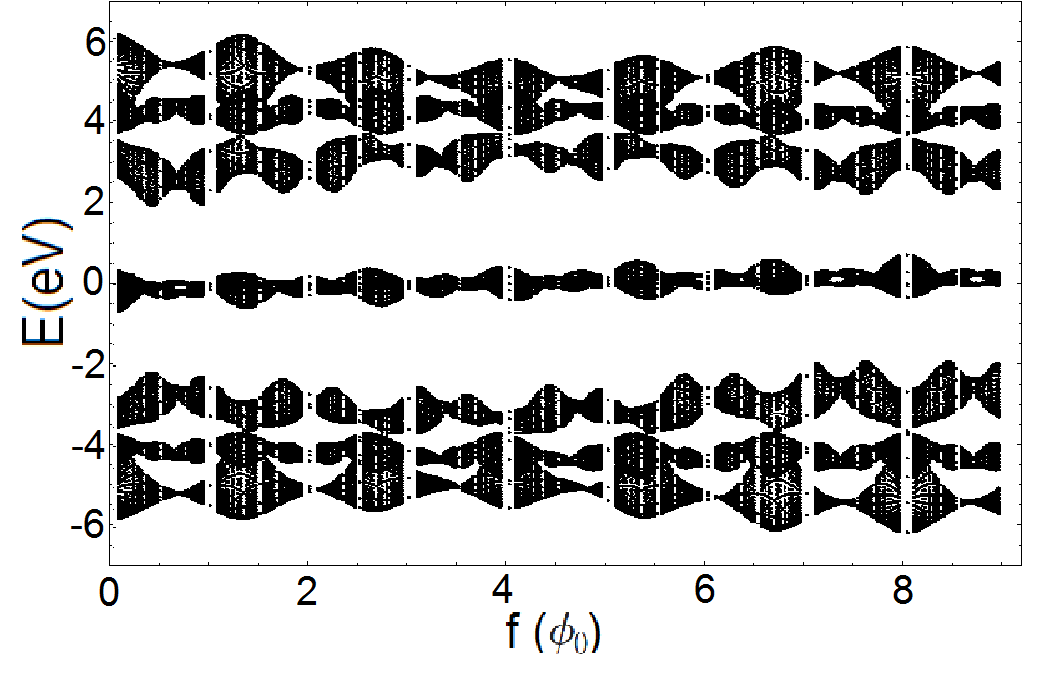}
		\caption{Electronic spectrum of the graphene (left) and the black phosphorene (right) nanoribbons depending on the magnetic flux.}\label{HB}
	\end{figure}

In Fig. \ref{HB}, the dependence of the electronic spectrum on the magnetic flux for both ZGNR's and ZPNR's is presented. Similarly to the analogous plots corresponding to the graphene monolayer in \cite{liu,wakabayashi2,QD}, the graphs in Fig. \ref{HB} show a fractal structure (called Hofst\"{a}dter butterfly \cite{hofstadter} in the case of the graphene monolayer). The emergence of the fractal structure is connected with the self-similarity of the electronic spectrum for different values of the magnetic flux \cite{wakabayashi2,FTO}. In the case of ZPNR, we see that the stability of the energy gaps is considerably strong not only against the changing size of the unit cell, but also against the magnetic field. Four new gaps emerge for a specific interval of the values of the magnetic flux. However, this is expectable if we check the character of DOS in the left part of Fig. \ref{vacDOS-ph}: the values in the energy intervals $-6\,{\rm eV}\,<E<-3.5\,{\rm eV}$ and $2.5\,{\rm eV}\,<E<6\,{\rm eV}$ are very close to zero in the case of the zero magnetic field.

We see that the magnetic field results in the emergence of new energy gaps in the electronic spectrum. This seems a bit unsatisfactory: the purpose of the addition of the magnetic field was the refinement of the undesirable effects of the edge vacancies on the metallic properties of ZGNR with a small unit cell, so we expect the vanishing of the energy gaps. But when we include the edge vacancies, we will see that both the emergence and the vanishing of the energy gaps occur. So we can expect an improvement of the metallic properties in some cases.\\

\subsection{Zigzag nanoribbons with edge vacancies}

\begin{figure}[htb]		
		 \includegraphics[width=\textwidth]{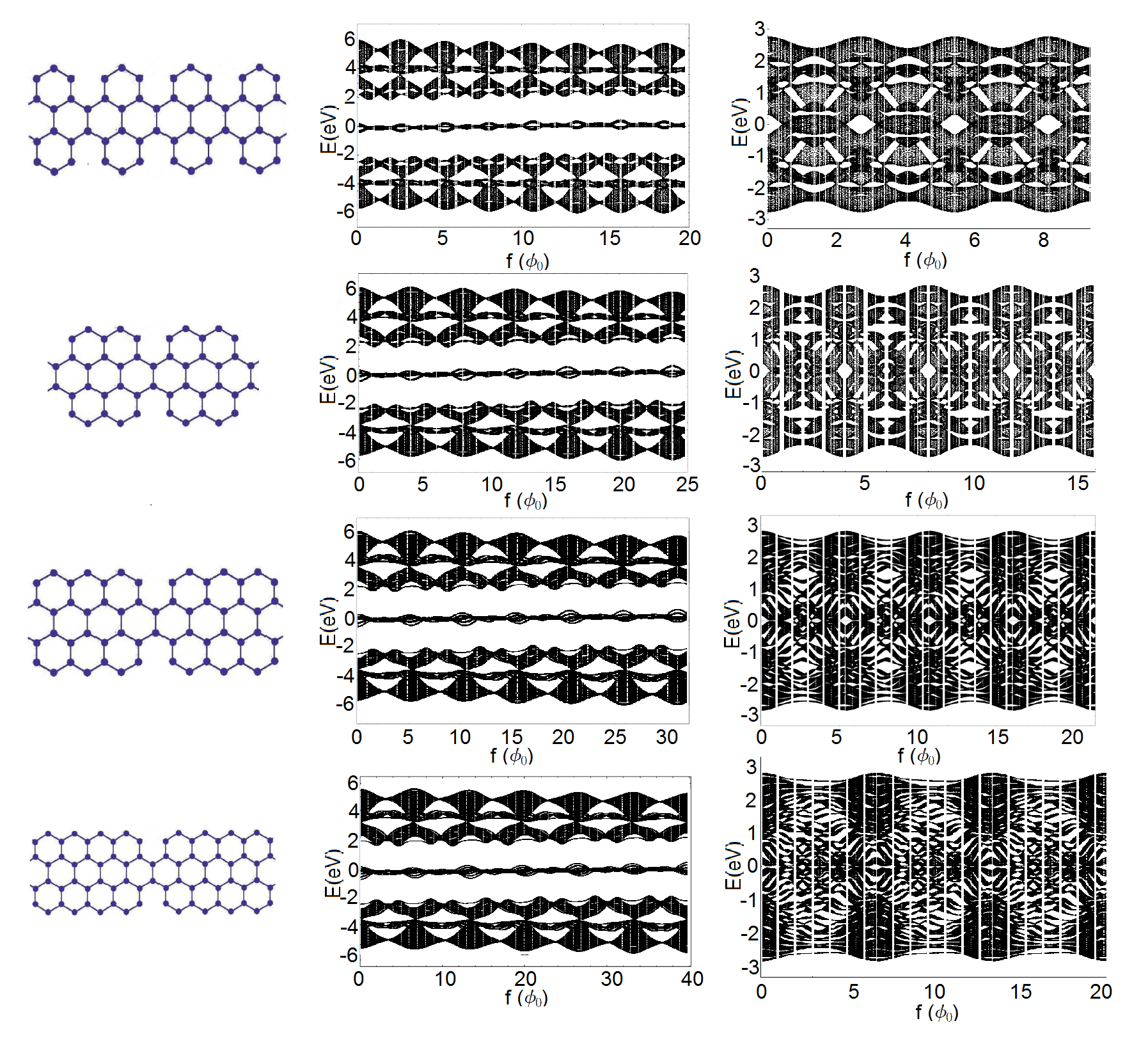}
		\caption{Electronic spectra depending on the magnetic flux for different kinds of ZPNR's (middle) and ZGNR's (right). The samples of the investigated nanostructures sketched in the left part, their (non-magnetic) unit cells correspond to those sketched in Fig. \ref{uc}.}\label{HB11}
\end{figure}

In Fig. \ref{HB11}, the dependence of the electronic spectrum on the magnetic flux is plotted for both ZPNR's and ZGNR's with different distances between 2 edge vacancies in the non-magnetic unit cell. The samples of the investigated nanoribbons with edge vacancies are plotted in the left part.

In the case of ZGNR's, we see that the magnetic field influences the width of the gaps in the electronic spectrum. Moreover, new gaps emerge (and other vanish). For each kind of ZGNR with a concrete distance between the edge vacancies in the unit cell, the placement of the gaps in the electronic spectrum is characteristic. For higher distances between the edge vacancies, the density of the gaps in the electronic spectrum is so high that for these structures (ZGNR's with the edge vacancies in the 3rd and the 4th line of Fig. \ref{HB11}), it would be worth investigating a possible emergence of new edge states for different values of the magnetic flux independently and comparing this effect with the results in \cite{turci}, where the addition of an impurity into the structure of the plain graphene causes the emergence of the localized states.

What is most important here: in all the studied types of ZGNR's, the gap at the Fermi level emerging in the absence of the magnetic field vanishes at a specific value of the magnetic flux, which enables the improvement of the metallic properties of ZGNR's with the edge vacancies in the case of a small unit cell (less than 13 atoms between 2 edge vacancies).
\begin{figure}[htb]		
		 \includegraphics[width=\textwidth]{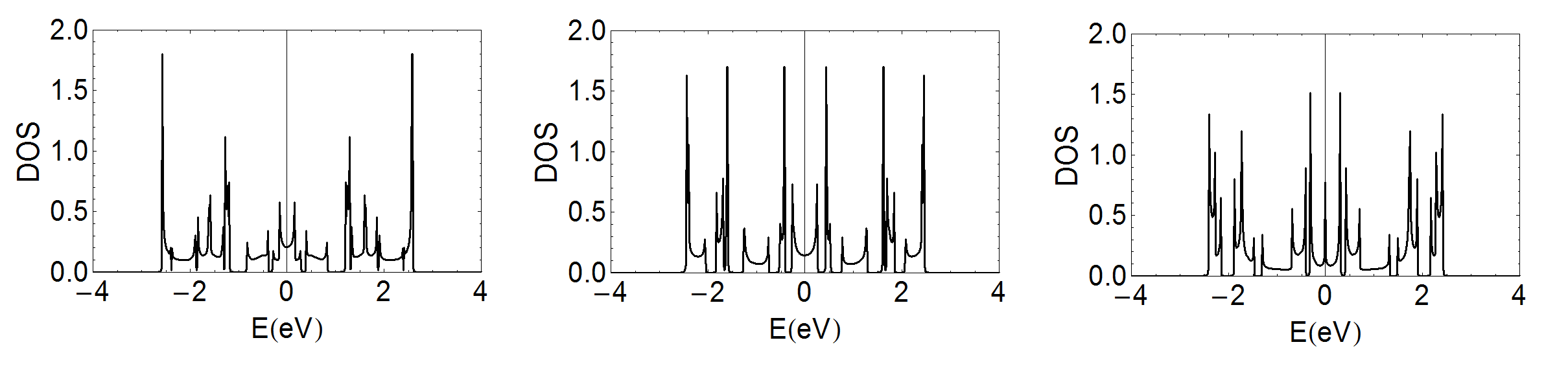}
		\caption{The density of states of ZGNR with vacancies sketched in first line of Fig. \ref{HB11} for different values of the magnetic flux: $f=2/3$ (left), $f=1$ (middle), $f=4/3$ (right).}\label{flux}
\end{figure}

In Fig. \ref{flux}, we see the rate of this possible improvement for ZGNR in the first line of Fig. \ref{HB11}. DOS is calculated here for 3 different values of the magnetic flux. As one can deduce, only the flux $f=4/3$ is sufficient to achieve a satisfactory reconstruction of the metallic properties of the corresponding nanostructure - the graph shows a peak for zero energy. In the other cases, there is a nonzero value of DOS for zero energy, so we can speak here about a weak metallization, but not a complete reconstruction of the metallic properties. So, we can say that for some unique values of the magnetic flux, the magnetic field can completely reconstruct the metallic properties of ZGNR with edge vacancies. But for most values of the magnetic flux, there is only a small improvement of these properties.

In the case of ZPNR's, the resulting spectrum does not differ very much from the case without the edge vacancies in Fig. \ref{HB}. The most visible change consists in the difference of the length of the period for different sizes of the unit cell. The position of the energy gaps is stable again. Moreover, we see the same effect which we observed in DOS in the middle and the right part of Fig. \ref{vacDOS-ph}: due to the stability of the energy gaps, a larger concentration of the energy states occurs in the places of the energy states corresponding to a shorter size of the unit cell.\\

\subsection{Conditions for the real occurrence of the predicted effects}\label{MSL}

All the studied effects connected with the switching on the uniform magnetic field face a big problem: a very strong magnetic field is needed to achieve the corresponding effects. It follows from the form of the magnetic phase factor mentioned at the beginning of this section: it has a form \cite{turci} \begin{equation}\gamma_{ij}\sim-\frac{\pi{\rm i}}{2}\frac{\Phi}{\Phi_0},\end{equation}
where
\begin{equation}\label{PhiB}\Phi=nB\cdot S,\hspace{1cm}\Phi_0=\hbar/e.\end{equation}
Here, $S=\frac{3}{2}a^2\sqrt{3}$ is the area of the hexagon, $n$ gives the number of the hexagons in the unit cell. This means that the magnetic factor depends on the length of the atomic bond $a$.

In the case of the large-area graphene or phosphorene, we can prove that an extremely large magnetic field is needed to achieve the emergence of the Hofst\"{a}dter butterfly in the electronic spectrum: let us suppose that $\Phi/\Phi_0=p/q=1/2$ and $n=4$ (typical size of the unit cell of the nanostructure with defects). Regarding the values of $\Phi_0$ and the length of the carbon--carbon bond, we can easily calculate that the required value of the magnetic field is $B\sim 1500$ T. Such a huge value reaches the outer limits of the laboratory experiments.

The situation would change if the lattice period given by the bond length $a$ would be enlarged significantly. Such a requirement is achieved in the moir\'{e} superlattice \cite{moire}, which is created when 1 atomic layer is overlaid over another nearly equivalent atomic layer -- there is a lattice mismatch between both layers. Then, the lattice period and the corresponding lattice constant increase many times. One of the possible examples is a graphene layer placed on the layer of the hexagonal boron nitride \cite{dean1}. Here, the lattice mismatch occurs which results in the lattice constant of the value 15 nm \cite{yang}; so the lattice constant increases approximately 100 times. As follows from (\ref{PhiB}), for a constant value of the magnetic flux $\Phi$, the magnetic field is inversely proportional to the second power of the lattice constant. So, in this case, the needed value of the magnetic field decreases approximately $10^4$ times, i.e., $B\sim 0.15$ T.

The introduced calculation is valid in the case of the large-area phosphorene or graphene. In the case of the nanoribbons, the situation differs because of the one-dimensional character of the corresponding nanostructure. Unlike the unit cells of the large-area planar superlattices, the unit cell of the moir\'{e} "supernanoribbon" has the area $k\cdot a\cdot a'$, where $k$ is a constant, $a$ is the bond length, and $a'$ is the lattice constant of the superlattice. It means that there is a linear dependence of the area of the unit cell on the lattice constant, and for a given value of the magnetic flux, the value of the magnetic field needed for the observation of the fractal structures in the energy spectrum (Fig. \ref{HB}) is inversely proportional to the first power of the lattice constant. This means that the 100-fold increase of the lattice constant ($1.42\,{\AA}\,\rightarrow\,15$ nm) causes the 100-fold decrease of the needed value of the magnetic field, i.e., $B\sim 15$ T. As it follows from \cite{dean1}, this range of values of the magnetic field is sufficient to observe the investigated effects.

According to \cite{dean1}, the Hofst\"{a}dter butterflies are observable if the magnetic length $l_B=\sqrt{\frac{\hbar}{eB}}$ is of the same order as the lattice constant. It can be derived from the previous expressions:
\begin{equation}a^2/l_B^2=\frac{2}{3n\sqrt{3}}\frac{\Phi}{\Phi_0}.\end{equation}
Using the values $\frac{\Phi}{\Phi_0}=1/2$, $n=4$, and $a=1.42\cdot 10^{-10}$ m , we get $a/l_B\sim 0.22$. So the orders of the lattice constant and of the magnetic length are comparable -- the fractal structure of the energy spectrum is observable in the studied cases (for the appropriate value of the magnetic field, i.e. 1500 T). \\

\section{Conclusion}

We studied the endurance of the metallic properties of ZPNR's and ZGNR's against the edge vacancies for the Gaussian distribution of these vacancies and different lengths of the unit cells and a possible influence of a uniform magnetic field on the metallic properties. Our results showed that except the case of a small size of the unit cell of ZGNR and some special configurations of the edge structure of the large unit cell, the metallic properties remained preserved.

For a small size of the unit cell, if we switch on a strong enough uniform magnetic field in the direction perpendicular to the surface, the gap at the Fermi energy in the electronic spectrum of ZGNR's vanishes for a specific value of the magnetic flux (Fig. \ref{HB11}, right). Furthermore, a lot of new gaps in the electronic spectrum emerge with such a high density that the emergence of new edge states could be expected. This will be one of the subjects of further investigations. The metallic properties are improved but they are completely restored for some extraordinary values of the magnetic flux only.

In the case of ZPNR's, the dependence of the electronic spectrum on the magnetic field has nearly the same character for an arbitrary size of the unit cell (Fig. \ref{HB11}, middle). A strong stability of the energy gaps in the electronic spectrum of ZPNR's emerges regardless of whether we enlarge the size of the unit cell or switch on the magnetic field: this causes that the density of the energy peaks in DOS increases in the places of the location of the energy peaks corresponding to DOS of the structure with a smaller size of the unit cell (Fig. \ref{vacDOS-ph}). As a consequence, the emergence of the new edge states could be expected here as well, but for zero magnetic field only.

The strong stability of the band gaps in the electronic spectrum of ZPNR's is a very interesting property. Its reason may be connected with the high number (5) of the hopping integrals. Due to them, the influence of the edge vacancies may not necessarily be sufficient to disturb most of the interconnections between the atoms. On the other hand, the hopping integrals $t_3, t_4$ and $t_5$ seem to be too weak in comparison with $t_1$ and $t_2$ to ensure the stable character of the electronic spectrum. The most likely explanation is a significant difference between the absolute values of $t_1$ and $t_2$: the ratio of them, $|t_1|/|t_2|=3$. In \cite{iran}, a considerable influence of this ratio on the form of the electronic spectrum and the probability amplitude is demonstrated. Other effects of this ratio will be the subject of the next investigations.

In summary, ZPNR's and ZGNR's are highly resistant against the edge vacancies. This makes them good candidates for the thermoelectric applications \cite{therm1,therm2,therm3,apl2}. Moreover, ZPNR's are suitable for the application as FET. Let us stress just only an illustrative character of our results: the edge vacancies are not the only kind of defects occurring in the process of fabrication, other defects may emerge inside the molecular structure. Moreover, the calculated values of the magnetic field needed for the improvement of the metallic properties in some cases are too high for us to do without further modifications, e.g., using moire superlattices (subsection \ref{MSL}). Other problems are connected with the stability of the zigzag edges: there were proposed some effective methods how to produce the corresponding nanoribbons \cite{ruff}, but their long-term storage is still a difficult problem. Due to the edge states, the zigzag edges show a high concentration of electrons, which results in a strong reactivity or in the transformation into a more stable "regzag" configuration \cite{koskinen}. So further investigation of the regzag configuration is one of the possibilities for the next calculations. In fact, this article is in some sense an extension of the above mentioned reference \cite{artem} which examines the semi-infinite graphene.\\

\section*{Acknowledgments}
The work was supported by VEGA Grant No. 2/0009/16. R. Pincak would like to thank the TH division at CERN for hospitality. Finally, we thank Prof. V. A. Osipov for valuable discussions, which helped us to improve the paper significantly.

\end{document}